\documentclass[twocolumn,aps,prc,superscriptaddress,reprint]{revtex4-1}
\makeatletter

\newcommand{\Rmnum}[1]{\expandafter\@slowromancap\romannumeral #1@}
\makeatother
\usepackage{blindtext}
\usepackage{graphicx}
\usepackage{dcolumn}
\usepackage{bm}
\usepackage{booktabs}
\usepackage{color}
\usepackage{mathrsfs}
\usepackage{ulem}
\usepackage{verbatim}
\usepackage{indentfirst}
\usepackage{float}
\usepackage{subfigure}
\usepackage{mathtools,nccmath}
\usepackage[colorlinks,
           bookmarks=true,
           linkcolor=blue,
           urlcolor=blue, 
            anchorcolor=black,
            citecolor=blue
            ]{hyperref}
\usepackage{hypcap}
\usepackage{amsmath}
\usepackage{epsfig}
\graphicspath{{plots/}}

\begin{document}
\title{Using local nuclear scaling of initial condition parameters to improve the system size dependence of transport model descriptions of nuclear collisions}

\author{Chao Zhang}
\affiliation{Key Laboratory of Quark and Lepton Physics (MOE) and Institute 
of Particle Physics, Central China Normal University, Wuhan 430079, China}
\affiliation{Department of Physics, East Carolina University, 
  Greenville, North Carolina 27858, USA} 
\author{Liang Zheng}
\affiliation{School of Mathematics and Physics, China University of
  Geosciences (Wuhan), Wuhan 430074, China}
\author{Shusu Shi} 
\affiliation{Key Laboratory of Quark and Lepton Physics (MOE) and Institute
of Particle Physics, Central China Normal University, Wuhan 430079, China}
\author{Zi-Wei Lin}\email{linz@ecu.edu}
\affiliation{Department of Physics, East Carolina University, 
  Greenville, North Carolina 27858, USA} 

\begin{abstract}
We extensively study the system size dependence of nuclear collisions 
with a multi-phase transport model. 
Previously certain key parameters for the initial condition needed significantly different values for $pp$ and central $AA$ collisions for the model to reasonably describe the yields and transverse momentum spectra of the bulk matter in those collision systems. 
Here we scale two key parameters, the Lund string fragmentation parameter $b_L$ and the minijet transverse  momentum cutoff $p_0$, with local nuclear thickness functions from the two colliding nuclei. 
This allows the model to use the parameter values for $pp$ collisions with the local nuclear scaling to describe the system size and centrality dependences of nuclear collisions self consistently. In addition to providing good descriptions of $pp$ collisions from 23.6 GeV to 13 TeV and reasonable descriptions of the centrality dependence of charged particle yields 
for Au+Au collisions from $7.7A$ GeV to $200A$ GeV and Pb+Pb collisions at LHC energies, 
the improved model can now well describe the centrality dependence of the mean transverse momentum of charged particles below $p_{\rm T} \lesssim 2$ GeV. It works similarly well for smaller systems including $p$Pb, Cu+Cu and Xe+Xe collisions.

\end{abstract}
\maketitle

\section{Introduction}

A main purpose of the field of high energy heavy ion collisions is to explore the properties of the produced hot and dense matter, the quark-gluon plasma (QGP). Many theoretical models   including transport models~\cite{Bass:1998ca,Xu:2004mz,Lin:2004en,Cassing:2009vt}, hydrodynamic models~\cite{Huovinen:2001cy,Betz:2008ka,Schenke:2010rr,Bozek:2011if}, and hybrid models~\cite{Petersen:2008dd,Werner:2010aa,Song:2010mg} are constructed to simulate and study the phase space evolution of the QGP. Comprehensive comparisons beween such models and the experimental data can provide us key information of the high density matter. 

In particular, the dependences of various observables on the size of the collision system or the centrality of a given collision system are useful as they may exhibit the onset or transition of certain phenomena such as the momentum anisotropy from initial state correlations~\cite{Dusling:2017dqg,Mace:2018vwq} or from final state interactions~\cite{He:2015hfa,Lin:2015ucn,Weller:2017tsr,Kurkela:2018ygx,Kurkela:2019kip}.
For large systems, it is commonly believed that viscous hydrodynamics applies well to the bulk of the matter, while transport model essentially approach the hydrodynamical limit since the average number of collisions per parton is large. 
For small colliding systems, however, hydrodynamic models and transport models may be quite different due to non equilibrium dynamics. 
Recently it has been found that parton transport can convert the initial spatial anisotropy into  significant anisotropic flows in the momentum space through the parton escape mechanism~\cite{He:2015hfa,Lin:2015ucn}, especially in small systems where the average number of collisions per particle is relatively small. Studies~\cite{Kurkela:2018ygx} also show that transport theory with a single scattering is very efficient in changing the particle distribution. Therefore, the system size dependence of anisotropic flows could provide key information on the origin of collectivity and the region of applicability of hydrodynamics in nuclear collisions. 

A multi-phase transport (AMPT) model, which we improve in this study, contains four main parts: the fluctuating initial condition from the HIJING  model~\cite{Wang:1991hta}, partonic interactions, hadronization, and hadronic interactions. 
The string melting version of AMPT model can reasonably describe many  
experimental data at low $p_{\rm T}$ in central and semi-central Au+Au collisions at $200A$ GeV and  Pb+Pb collisions at the LHC~\cite{Lin:2014tya,He:2017tla} including the pion, kaon, proton yields, $p_{\rm T}$ spectra and elliptic flow. 
Recently we updated the AMPT model with a new quark coalescence model~\cite{He:2017tla} and modern nuclear parton distribution functions (nPDFs)~\cite{Zhang:2019utb}, 
where the string melting version can also reasonably describe the charged particle rapidity distributions and $p_{\rm T}$ spectra in $pp$ collisions at different energies. 
On the other hand, from the comparison to experimental data 
we have found that certain key parameters in the AMPT model 
need to have very different values for $pp$ and central $AA$ collisions. 
First, the $b$ parameter in the Lund symmetric fragmentation function
~\cite{Andersson:1983jt,Andersson:1983ia} (denoted as $b_L$ in this study) 
needs to be $\approx 0.15/$GeV$^{-2}$ for central Au+Au or Pb+Pb collisions, which is a few times smaller than its value for $pp$ collisions. 
Second, the minijet transverse momentum cutoff $p_0$ for central  Pb+Pb collisions at the LHC energies needs to be significantly bigger than its value for $pp$ collisions at the same energy in the AMPT model updated with modern nPDFs ~\cite{Zhang:2019utb}.  
These observations clearly indicate that these two parameters should depend on the size of the colliding system.

In this study we improve the system size and centrality dependences of the AMPT model~\cite{Lin:2004en} by treating the two parameters in its initial condition, the Lund $b_L$ parameter and the minijet cutoff $p_0$, as local variables that depend on the local nuclear thickness functions, $T_A(s_A)$ and $T_B(s_B)$, from the two colliding nuclei in each event. The rest of the paper is organized as follows. In Sec.~\ref{sec:improve} we discuss the local nuclear scaling of the Lund $b_L$ and momentum cutoff $p_0$ parameters. We then 
systematically compare results from the improved AMPT model 
with the experimental data for $pp$, $pA$, and $AA$ collisions at RHIC and LHC energies in Sec.~\ref{sec:results}, including the charged particle yields and $p_{\rm T}$ spectrum or mean $p_{\rm T}$ and their centrality dependences in nuclear collisions. 
After more discussions in Sec.~\ref{sec:discussions}, we then summarize in Sec.~\ref{sec:summary}.

\section{improvement of the initial condition of the AMPT model}
\label{sec:improve} 
The initial condition component of the AMPT model is based on the HIJING two component model~\cite{Wang:1991hta}. 
The primary interactions between the two incoming nuclei are divided into two components:  the soft component  described by the Lund string fragmentation model~\cite{Andersson:1983jt,Andersson:1983ia,Sjostrand:1993yb} that includes the parameter $b_L$, and the hard component with a minijet transverse momentum cutoff $p_0$ that is described by perturbative QCD through the PYTHIA program~\cite{Sjostrand:1993yb}.
Rather than treating $p_0$ and $b_L$ as constant parameters (at least for a given collision system at a given energy), as done in almost all previous studies with the AMPT model, here we model them as local variables that depend on the nuclear thickness functions of the two nuclei.

\subsection{Local Lund string fragmentation parameter $b_L$}

In the Lund string model~\cite{Andersson:1983jt,Andersson:1983ia}, 
the symmetric fragmentation function is given by 
\begin{align}
f(z) \propto z^{-1}(1-z)^{a_L}~e^{-b_L m^2_{\rm T}/z}, 
\end{align} 
where $z$ is the light-cone momentum fraction of the 
produced hadron with respect to the fragmenting string, and 
$m_{\rm T}$ is the hadron transverse mass. 
The average squared transverse momentum of massless hadrons 
from fragmentation is then related to the Lund fragmentation parameters $a_L$ and $b_L$ as~\cite{Lin:2004en}
\begin{align}
\label{eq:pt2}
\langle p_{\rm T}^{2}\rangle=\frac{1}{b_L(2+a_L)}.
\end{align}
As a result, the average $p_{\rm T}$ of partons after string melting 
and consequently the final hadron $p_{\rm T}$ spectrum from the string melting version of the AMPT model strongly  depend on $b_L$. In the AMPT model updated with modern nPDFs~\cite{Zhang:2019utb}, we have found that a constant values of $a_L=0.8$ and $b_L=0.4$ GeV$^{-2}$ can reasonably describe the $p_{\rm T}$ spectra of $pp$ and $p\bar{p}$ collisions over a wide energy range. 
However, a much smaller value of $b_L \approx 0.15$ GeV$^{-2}$ is needed to describe the $p_{\rm T}$ spectra in central Au+Au collisions at RHIC and central Pb+Pb collisions at LHC~\cite{Lin:2014tya,He:2017tla,Zhang:2019utb}. 
It was also realized that the centrality dependence of the charged particle mean transverse momentum $\langle p_{\rm T} \rangle$ in heavy ion collisions has the opposite trend in  comparison to the experimental data~\cite{Ma:2016fve}, where the system size dependence of the Lund fragmentation parameters was proposed as a possible solution.
Since we expect the mean transverse momentum of initial partons to be higher in larger systems due to the higher initial temperature and Eq.\eqref{eq:pt2} relates the mean squared transverse momentum after string melting to the Lund $b_L$ parameter,  
it is natural that $b_L$ should depend on the system size.

We now make $b_L$ a local variable that depends on the transverse position of the corresponding excited string in each event. Note that its value has been found to be smaller for a larger collision system, which is consistent with the expectation of a stronger color field and thus a higher string tension $\kappa$ since $\kappa \propto 1/b_L$\cite{Lin:2004en}. 
Therefore we scale $b_L$ with the local nuclear thickness functions in a general $AB$ collision as \begin{eqnarray}
\label{eq:bl}
b_L(s_A,s_B,s)=\frac {b_L^{pp}} {\left [ \sqrt {T_A(s_A) T_B(s_B)}/T_p \right ]^{\beta(s)}}.
\end{eqnarray}
In the above, 
$b_L^{pp}$ is the value for $pp$ collisions (to be discussed further in Sec.~\ref{subsec:pp}), 
$s$ is the square of the center-of-mass collision energy per nucleon pair, 
$T_A(s_A)=\int\rho_{A}(s_A,z)dz$ is the nuclear thickness function 
at the transverse distance $s_A$ from the center of nucleus $A$ from Woods-Saxon nuclear  density profiles~\cite{Eskola:1998iy}, 
and $T_p$ (taking the value of $0.22$ fm$^{-2}$ in this study) can be considered as the average value of the effective thickness function of the proton. 
Note that in Eq.\eqref{eq:bl} (and Eqs. \eqref{eq:p0}, \eqref{eq:blplus} and \eqref{eq:p0plus}) 
$T_p$ is used instead of $T_A(s_A)$ or $T_B(s_B)$ when the projectile or the target is proton or when $T_A(s_A)$ or $T_B(s_B)$ from the nucleus is smaller than the $T_p$ value.
Also note that there are two types of strings in the fragmentation process. 
The first type is a wounded nucleon from the projectile (or target) nucleus that has interacted with one or more nucleons in the target (or projectile);  
we take the nucleon position in the nucleus $s_A$ (or $s_B$) in Eq.\eqref{eq:bl} 
and then for simplicity take the other position $s_B$ (or $s_A$) via the relation 
$\vec{s_B}=\vec{s_A}+\vec{b}$. 
The other type is an independent string from the primary nucleon-nucleon interaction through the hard process, where the values of both $s_A$ and $s_B$ are unique 
and thus directly used in Eq.\eqref{eq:bl}.

\begin{figure}[!htb]
\includegraphics[scale=0.43]{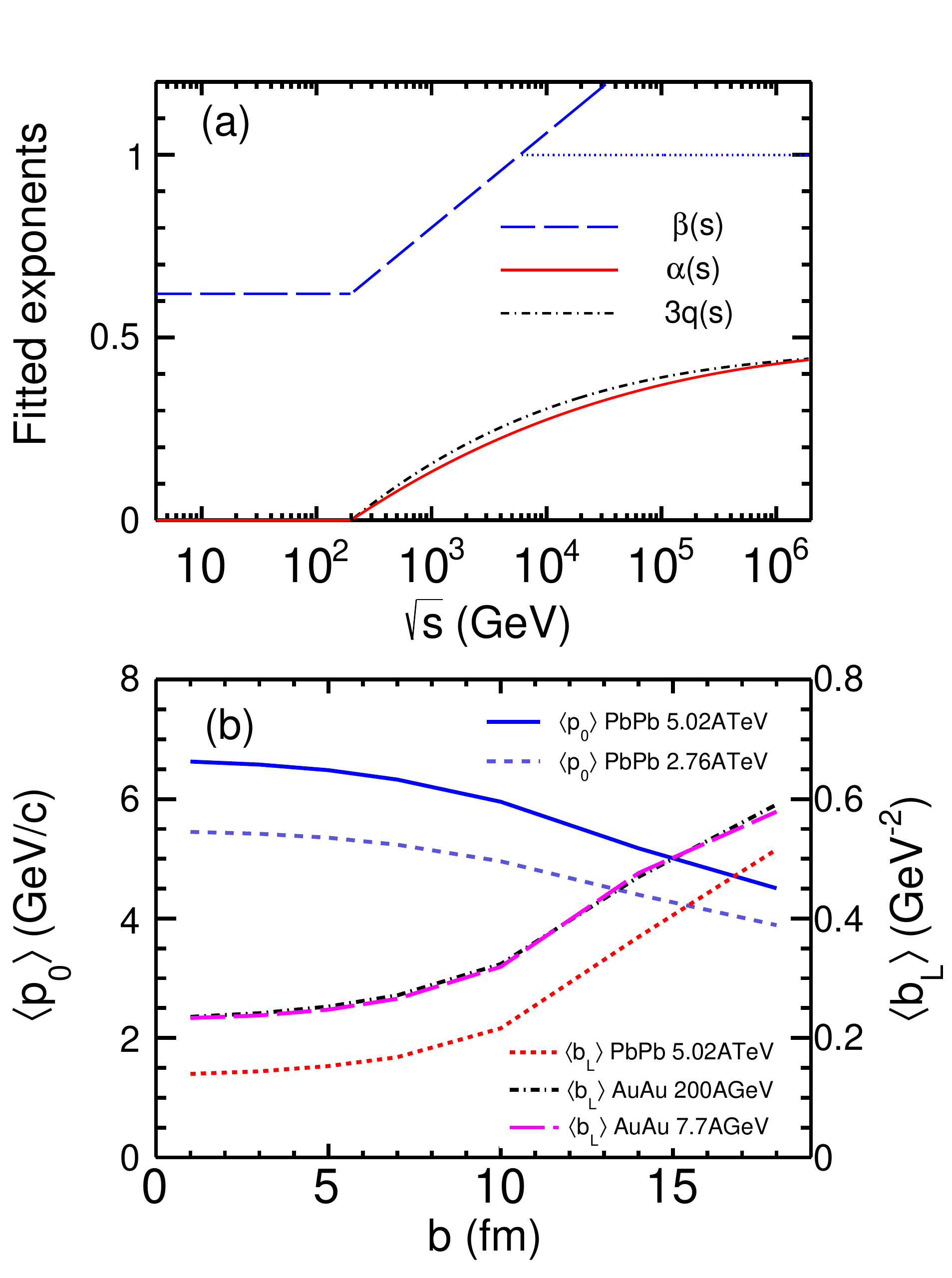}
\caption{(a) Fitted exponent functions $\beta(s)$ and $\alpha(s)$ versus the 
center-of-mass energy per nucleon pair $\sqrt{s}$, where the function $3q(s)$ is also shown for comparison. 
(b) Average $p_0$ and $b_L$ values versus the impact parameter in Pb+Pb and Au+Au collisions at several energies.}
\label{fig:blp0ab}
\end{figure}

First we have determined that a constant value $b_L^{pp}=0.7$ GeV$^{-2}$ provides a reasonably good description of the charged particle $\langle p_{\rm T} \rangle$ in $pp$ collisions (details in Sec.\ref{subsec:pp}). Next we fit the charged particle $\langle p_{\rm T} \rangle$ in the most central Au+Au collisions at RHIC energies and most central Pb+Pb collisions at LHC energies to obtain the preferred $\beta(s)$ value at each of those energies. 
The results show that the preferred $\beta(s)$ is almost a constant at RHIC energies but 
needs to be significantly bigger at LHC energies. 
We thus parametrize the $\beta(s)$ function as
\begin{eqnarray}
\beta(s)=0.620+ 0.112\ln \left (\frac{\sqrt{s}}{E_0} \right ) \Theta (\sqrt{s}-E_0),
\end{eqnarray}
where $E_0=200$ GeV, $\sqrt{s}$ is the center-of-mass collision energy per nucleon pair, 
and $\Theta(x)$ is the theta function. 
The fitted $\beta(s)$ function is shown in Fig.~\ref{fig:blp0ab}(a) (dashed line). 
Note that the fit is not constrained or tested by data between the energy of $200A$ GeV and $2.76A$ TeV or above $5.44A$ TeV due to the lack of heavy ion data. On the other hand, 
the value of $\beta=1$ (dotted line) may be a ``natural'' limit for Eq.\eqref{eq:bl} at high energies if all local strings would fully overlap so that the string tension would add up, since it corresponds to $b_L  \propto 1/T_A(s_A)$ for central $AA$ collisions where $T_A(s_A)$ is proportional to the local number of participant nucleons or excited strings integrated over the longitudinal length.

\begin{figure}[!htb]
\includegraphics[scale=0.43]{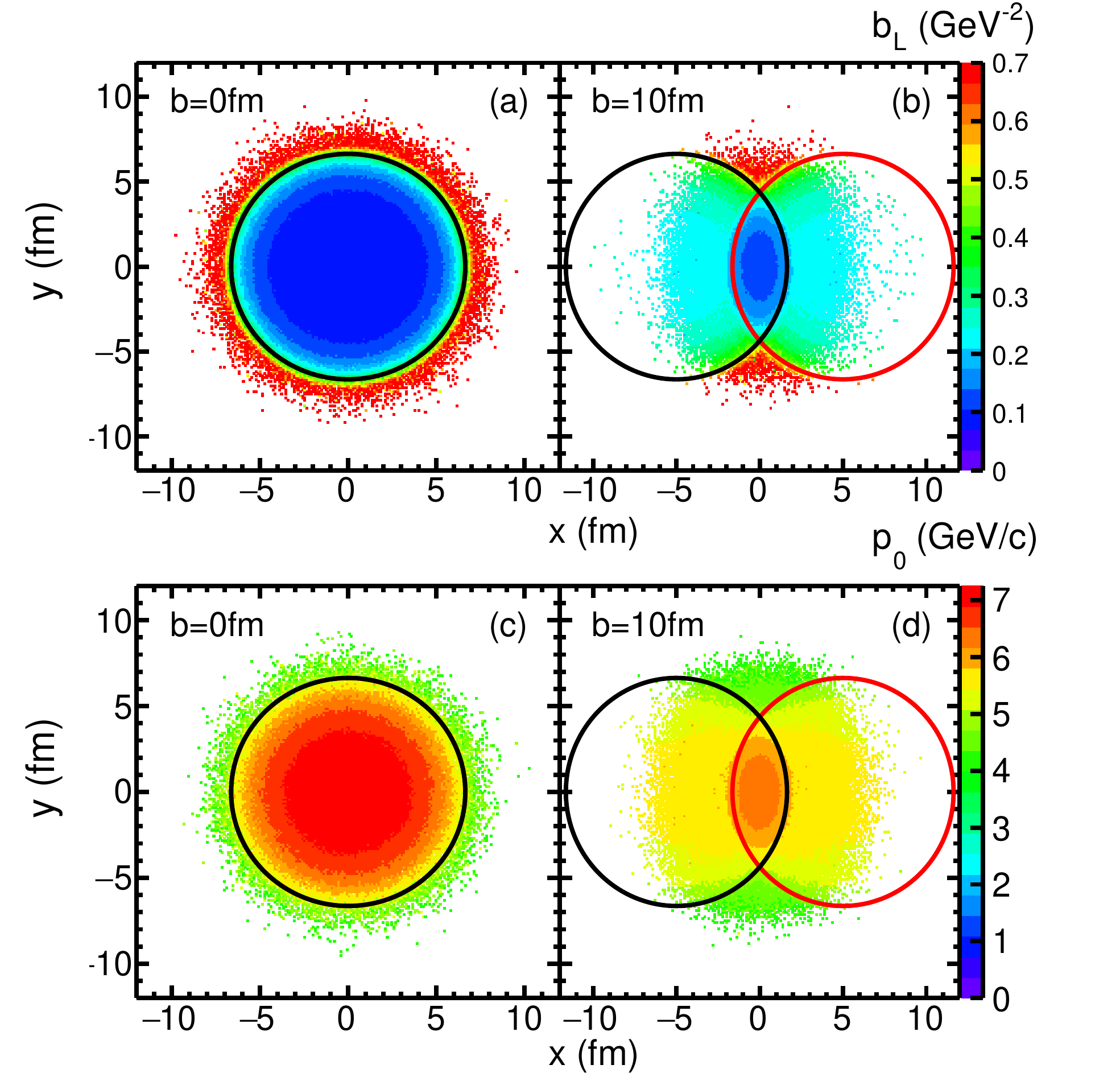}
\caption{Distributions of $b_L$ values of Eq.\eqref{eq:bl} (upper panels) and $p_0$ values of Eq.\eqref{eq:p0} (lower panels) over the transverse plane from multiple central (left panels) and peripheral (right panels) Pb+Pb  events at $5.02A$ TeV from the AMPT model.}
\label{fig:blp0}
\end{figure}

Figures~\ref{fig:blp0}(a) and \ref{fig:blp0}(b) show the distributions of $b_L$ values of  Eq.\eqref{eq:bl} over the transverse plane of multiple central ($b=0$) and peripheral ($b=10$ fm) $5.02A$ TeV Pb+Pb events, respectively, from the AMPT model simulations. 
Specifically, each point represents the $b_L$ value of a wounded nucleon 
or an independent string versus its transverse position in the collision.
The red and black circles represent the hard-sphere boundaries of the projectile and target nuclei, respectively, to indicate the scale. 
We see that the values in less-overlapped regions are close to the value for $pp$ collisions, while the $b_L$ values in highly-overlapped regions are much lower. 
Figure~\ref{fig:blp0ab}(b) shows the $b_L$ value averaged over the overlap volume 
as a function of the impact parameter for Pb+Pb collisions at $5.02A$ TeV and Au+Au collisions at two RHIC energies.
We see that as expected $\langle b_L \rangle$ at the LHC energy is lower due to the larger value of the exponent $\beta(s)$, while the impact parameter dependences of $\langle b_L \rangle$ at different RHIC energies are essentially the same due to the constant value of $\beta(s)$ within that energy range.

\subsection{Local minijet transverse momentum cutoff $p_0$}
In the hard component of the HIJING model, the total jet cross section is given by 
\begin{align}
\label{eq:sigjet}
\sigma_{\rm jet}=\sum_{c,d}\frac{1}{1+\delta_{cd}}\int dy_1dy_2 \int_{p_0^2}^{\hat s/4}dp_{\rm T}^2\frac{d\sigma^{cd}}{dp_{\rm T}^{2}dy_1dy_2}.
\end{align}
In the above, $p_0$ is the cutoff of the minijet transverse momentum, 
$\hat s$ is the Mandelstam variable for the minijet production subprocess, 
and $d\sigma^{cd}/(dp_{\rm T}^{2}dy_1dy_2)$ is the differential cross section ~\cite{Eichten:1984eu} for the two colliding nuclei to produce the pair of minijet partons of flavor $c$ and $d$ at  rapidity $y_1$ and $y_2$, respectively. The $p_0$ cutoff (relevant when $\sqrt{s}>10$ GeV) 
together with the soft component cross section ($\sigma_{\rm soft}$) are the two key parameters to determine the total, elastic and inelastic cross sections of nuclear collisions in the HIJING  model~\cite{Wang:1991hta,Deng:2010mv,Zhang:2019utb}, 
In our recent update of the AMPT model with modern nPDFs~\cite{Zhang:2019utb}, 
using the $pp$ cross section data we determined 
$p_0$ and $\sigma_{\rm soft}$ as functions of the colliding energy. 
Motivated by the physics of color glass condensate~\cite{McLerran:1993ni}, 
we further introduced a nuclear scaling of the $p_0$ cutoff 
for central $AA$ collisions above the top RHIC energy of $200A$ GeV 
to describe the experimental data on charged particle yields in central Pb+Pb collisions at LHC energies. 
That scaling~\cite{Zhang:2019utb} can be considered as a global nuclear scaling because the scaled $p_0$ value is a constant for all events of a given collision system at a given energy.

In a subsequent work that improved heavy flavor productions~\cite{Zheng:2019alz}, 
we started to use the minijet cross section as shown in Eq.\eqref{eq:sigjet}, 
which changed the factor of 1/2 in the original HIJING model~\cite{Wang:1991hta,Lin:2004en,Zhang:2019utb} to 
$1/(1+\delta_{cd})$ to differentiate minijet final states with or without identical partons. 
We also removed the momentum cutoff for heavy quark productions and then 
included heavy quark production cross sections in the total minijet cross section~\cite{Zheng:2019alz}. 
These modifications have little effect on $\sigma_{\rm soft}$, but they lead to an increase of the total minijet cross section and consequently a small increase of the $p_0$ cutoff in $pp$ collisions as given below:
\begin{eqnarray}
\label{eq:p0pp}
p_0^{pp}(s)&=&-1.92 +  1.77\ln(\sqrt{s})-0.274\ln^{2}(\sqrt{s}) \nonumber \\ 
&&+0.0176\ln^{3}(\sqrt{s})
\end{eqnarray}
with $\sqrt{s}$ in GeV. 
For the global nuclear scaling relation $p_0^{AA}(s)=p_0^{pp}(s) A^{q(s)}$~\cite{Zheng:2019alz}, the modifications also lead to a small change of the nuclear scaling exponent $q(s)$ for central $AA$ collisions:
$q(s)=0.0369\ln (\sqrt{s}/E_0)-0.00318\ln^2(\sqrt{s}/E_0) +0.0000990\ln^3(\sqrt{s}/E_0)$ for $\sqrt {s} \geq E_0$ while $q(s)=0$ for $\sqrt {s}< E_0$. 
On the other hand, we do not expect the global nuclear scaling to hold for non-central $AA$ collisions; for example, we expect little nuclear scaling for very peripheral $AA$ collisions since they should be similar to $pp$ collisions. Indeed, we have shown~\cite{Zhang:2019utb} that the charged particle yield in peripheral Pb+Pb collisions at $5.02A$ TeV is better described without using the global nuclear scaling of $p_0$, although the scaling is necessary for central Pb+Pb collisions.

We now go beyond the global nuclear scaling and instead make $p_0$ a local variable that depends on the transverse position of the corresponding hard process in each event. 
Since the $p_0$ cutoff has been found to increase with the system size, it is natural to relate it to the nuclear thickness functions in a general $AB$ collision; thus we write
\begin{eqnarray}
\label{eq:p0}
p_0(s_A,s_B,s)&=&p_0^{pp}(s) * \left [ \sqrt {T_A(s_A) T_B(s_B)}/T_p \right ]^{\alpha(s)}.
\end{eqnarray}
Since $T_A(s_A) \propto A^{1/3}$, Eq.\eqref{eq:p0} approximately gives 
$p_0 \propto A^{\alpha(s)/3}$ for central $AA$ collisions and thus essentially recovers the previous global nuclear scaling relation if $\alpha(s)=3q(s)$.
On the other hand, for peripheral collisions $T_A(s_A)$ and $T_B(s_B)$ are expected to be small and close to the proton value ($T_p$), then Eq.\eqref{eq:p0} automatically gives   
the $p_0$ value for $pp$ collisions. This way Eq.\eqref{eq:p0} captures the expected system size dependence as well the centrality dependence of the $p_0$ cutoff parameter. 

From the comparison to charged particle yields in the most central Pb+Pb collisions at $2.76A$ TeV and $5.02A$ TeV, we obtain the preferred $\alpha(s)$ values at those two energies. 
Since $p_0^{pp}(s)$ works for central Au+Au collisions at $200A$ GeV, we assume that the need to modify $p_0$ in nuclear collisions starts at the top RHIC energy~\cite{Zhang:2019utb}. We then fit the $\alpha(s)$ function as
\begin{eqnarray}
\alpha(s)&=&0.0918\ln \left (\frac{\sqrt{s}}{E_0} \right )- 0.00602
\ln^{2}  \left ( \frac{\sqrt{s}}{E_0} \right ) \nonumber \\
&&+0.000134 \ln^{3}  \left ( \frac{\sqrt{s}}{E_0} \right ), {\rm ~for~} \sqrt{s} \geq E_0 
\end{eqnarray}
with $\alpha(s)=0$ for $\sqrt{s} < E_0$. 
As shown in Fig.~\ref{fig:blp0ab}(a), $\alpha(s) \approx 3q(s)$ as expected, 
and both have values close to 1/2 at the very high energy of $10^8$ GeV. Note that the high energy $q(s)$ value of about 1/6~\cite{Zhang:2019utb} is motivated by the color glass condensate~\cite{McLerran:1993ni}, 
where the saturation momentum $Q_s$ scales with the nuclear size as $Q_s \propto A^{1/6}$ in the saturation regime.

Figures~\ref{fig:blp0}(c) and \ref{fig:blp0}(d) show the distributions of $p_0$ values of  Eq.\eqref{eq:p0} over the transverse plane of multiple central ($b=0$) and peripheral ($b=10$ fm) simulated $5.02A$ TeV Pb+Pb events, respectively. 
Each point represents the $p_0$ value of a wounded nucleon that is involved in hard processes versus its transverse position in the collision. 
Similar to Figs.~\ref{fig:blp0}(a) and \ref{fig:blp0}(b), we see that the $p_0$ value varies from $p_0^{pp}$ ($\approx 4.2$ GeV at this energy) in  less-overlapped regions to bigger values in highly-overlapped regions as expected, and the variation is larger for more central collisions. 
In addition, the relative variation of the $p_0$ values is much smaller than that of the $b_L$ values because $\alpha(s) \ll \beta(s)$ for the exponents. 
The average $p_0$ value, i.e., averaged over the wounded nucleons in the overlap volume, is shown in Fig.~\ref{fig:blp0ab}(b) as a function of the impact parameter for Pb+Pb collisions at $2.76A$ TeV and $5.02A$ TeV. 
We see that $\langle p_0 \rangle$ gradually decreases with the increase of impact parameter and that $\langle p_0 \rangle$ is smaller at the lower LHC energy due to the smaller $\alpha(s)$ value there.

\section{results for various collision systems}
\label{sec:results}

We apply the local Lund parameter $b_L$ of Eq.\eqref{eq:bl} and local minijet cutoff $p_0$ of Eq.\eqref{eq:p0} to systematically study charged particle productions in different collision systems over a wide range of energies.

\subsection{$pp$ and $p\bar{p}$ collisions}
\label{subsec:pp}

\begin{figure}[!htb]
\includegraphics[scale=0.43]{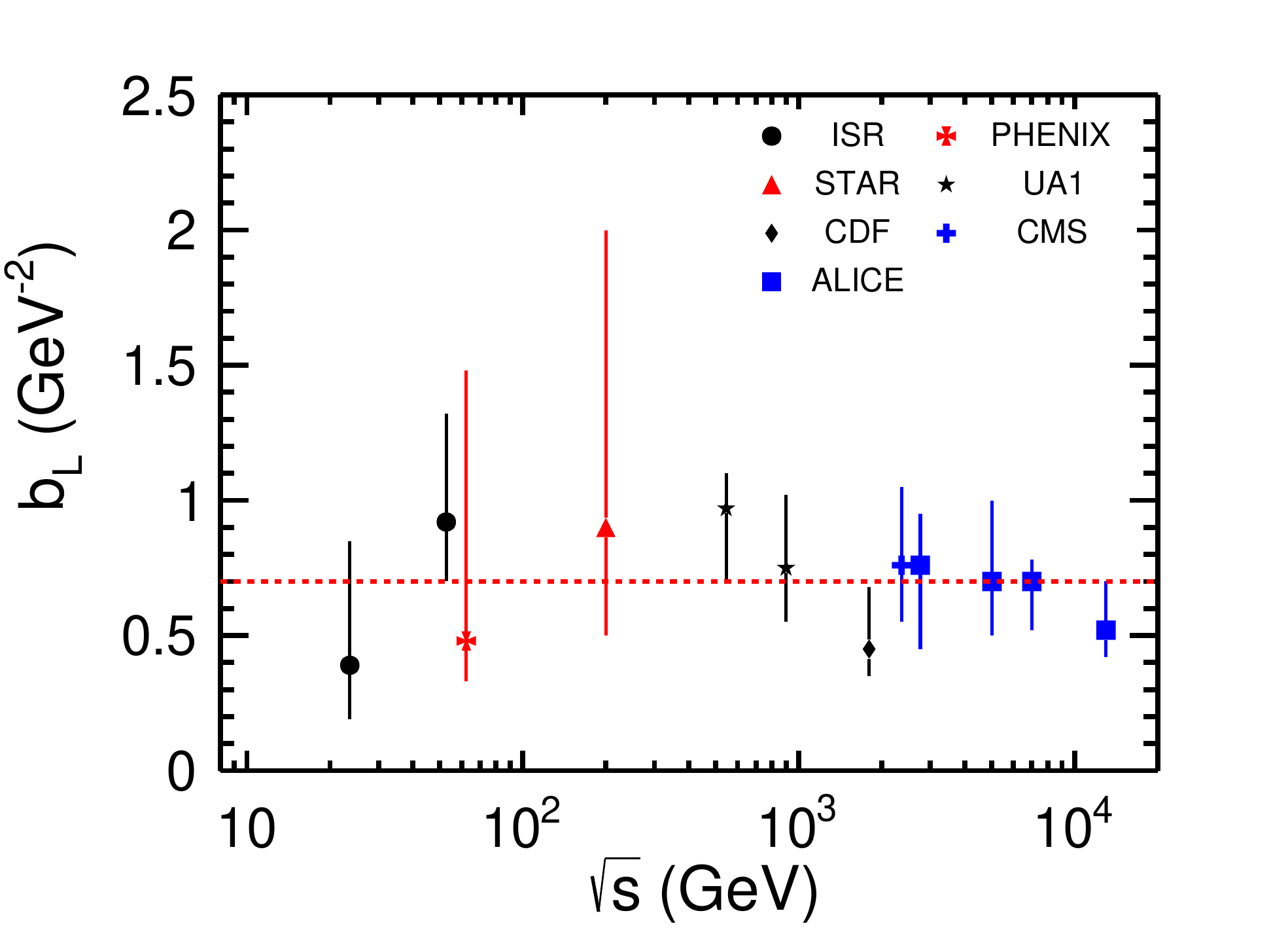}
\caption{The preferred individual $b_L$ values from fitting the experimental $\langle p_{\rm T} \rangle$ of charged particles in $pp$ or $p\bar{p}$ collisions at different energies; 
the error bar corresponds to an assumed 3\% uncertainty of the experimental $\langle p_{\rm T}\rangle$ value, while the dashed line represents our choice of $b_L^{pp}=0.7$ GeV$^{-2}$ in this study.}
\label{fig:blpp}
\end{figure}

For $pp$ and $p\bar{p}$ collisions we first determine the value of $b_L^{pp}$ in Eq.\eqref{eq:bl}, while the minijet cutoff $p_0^{pp}(s)$ has been specified in Eq.\eqref{eq:p0pp}. 
We first obtain the preferred value of $b_L$ at each energy (symbols in Fig.~\ref{fig:blpp}) by fitting the mean $p_{\rm T}$ data of charged particles in $pp$ or $p\bar{p}$ collisions from 23.6 GeV to 13 TeV ($p\bar{p}$ at 546, 900, and 1800 GeV). 
The uncertainty of $b_L$ at each energy is obtained by assuming a 3\% uncertainty for the experimental $\langle p_{\rm T}\rangle$ value.
We see that the preferred central value of $b_L$ fluctuates approximately within [0.4-1.0] GeV$^{-2}$ and a constant value of 0.7 GeV$^{-2}$ (dashed line) describes the experimental $\langle p_{\rm T}\rangle$ data within about 3\%. Therefore we take $b_L^{pp}=0.7$ GeV$^{-2}$ for $pp$ and $p\bar{p}$ collisions at all energies. 
Note that a constant Lund parameter $a_L=0.8$ is taken for all collision systems at all energies in the AMPT model improved with modern nPDFs~\cite{Zhang:2019utb,Zheng:2019alz}. 

The $\langle p_{\rm T}\rangle$ in this study is calculated for charged hadrons up to $p_{\rm T} \approx 2$ GeV$/c$ for both the AMPT results and the experimental data, because the AMPT model cannot reliably be  used for high $p_{\rm T}$ hadrons due to its lack of the radiative energy loss and independent fragmentation of high $p_{\rm T}$ partons. 
Note the different $p_{\rm T}$ ranges used for the $\langle p_{\rm T}\rangle$ calculation in Fig.~\ref{fig:blpp}: 
[0.90, 2.15] GeV$/c$ at 23.6 GeV~\cite{Thome:1977ky}, [0.73, 2.10] GeV$/c$ at 53 GeV~\cite{Thome:1977ky}, [0.6, 2] GeV$/c$ at 62.4 GeV~\cite{Adare:2011vy}, 
[0.2, 2] GeV$/c$ at 200 GeV~\cite{Adams:2003kv}, 546 GeV and 900 GeV~\cite{Albajar:1989an}, [0.1, 2] GeV$/c$ at 2.36 TeV~\cite{Roland:2010ema}, and [0.15, 2] GeV$/c$ at 1.8 TeV~\cite{Abe:1989td}, 2.76 TeV~\cite{Abelev:2013ala}, 
5.02 TeV~\cite{Acharya:2018qsh}, 7 TeV~\cite{Khachatryan:2010us} and 13TeV~\cite{Adam:2015pza}.
Also note that in this study we treat charged particles from the AMPT model more carefully in the comparisons with data. Specifically, we decay the $\Sigma^{+}, \Sigma^{-}$ 
hyperons including their antiparticles as well as all open charm hadrons 
(with PYTHIA~\cite{Sjostrand:2006za}) before calculating charged particle observables. 
This treatment leads to a slight increase of the charged particle yield at low $p_{\rm T}$ but a slight decrease at high $p_{\rm T}$ (by several percent) compared to results using the  previous analysis method~\cite{Zhang:2019utb}. 

\begin{figure}[!htb]
\includegraphics[scale=0.40]{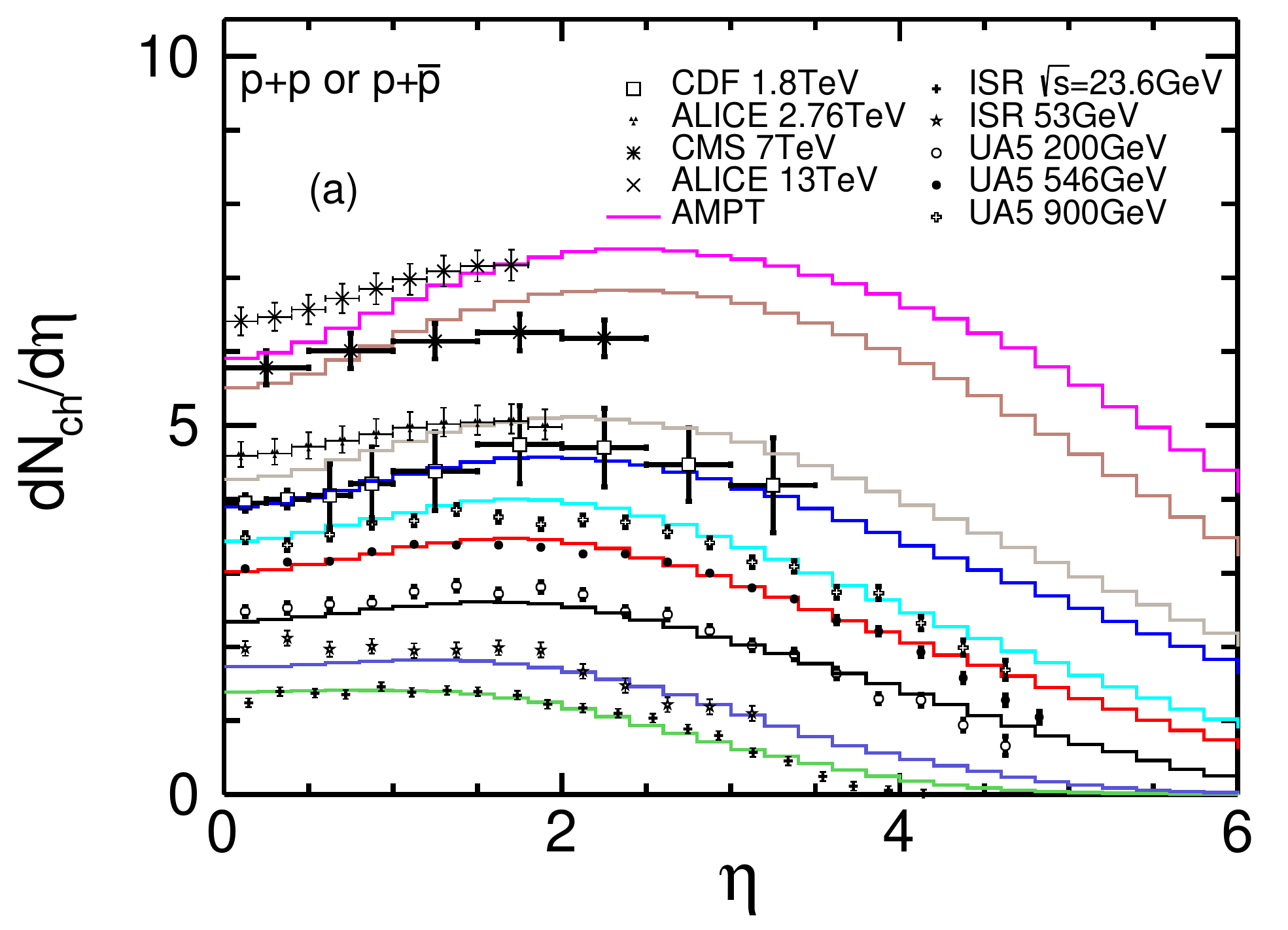}
\includegraphics[scale=0.40]{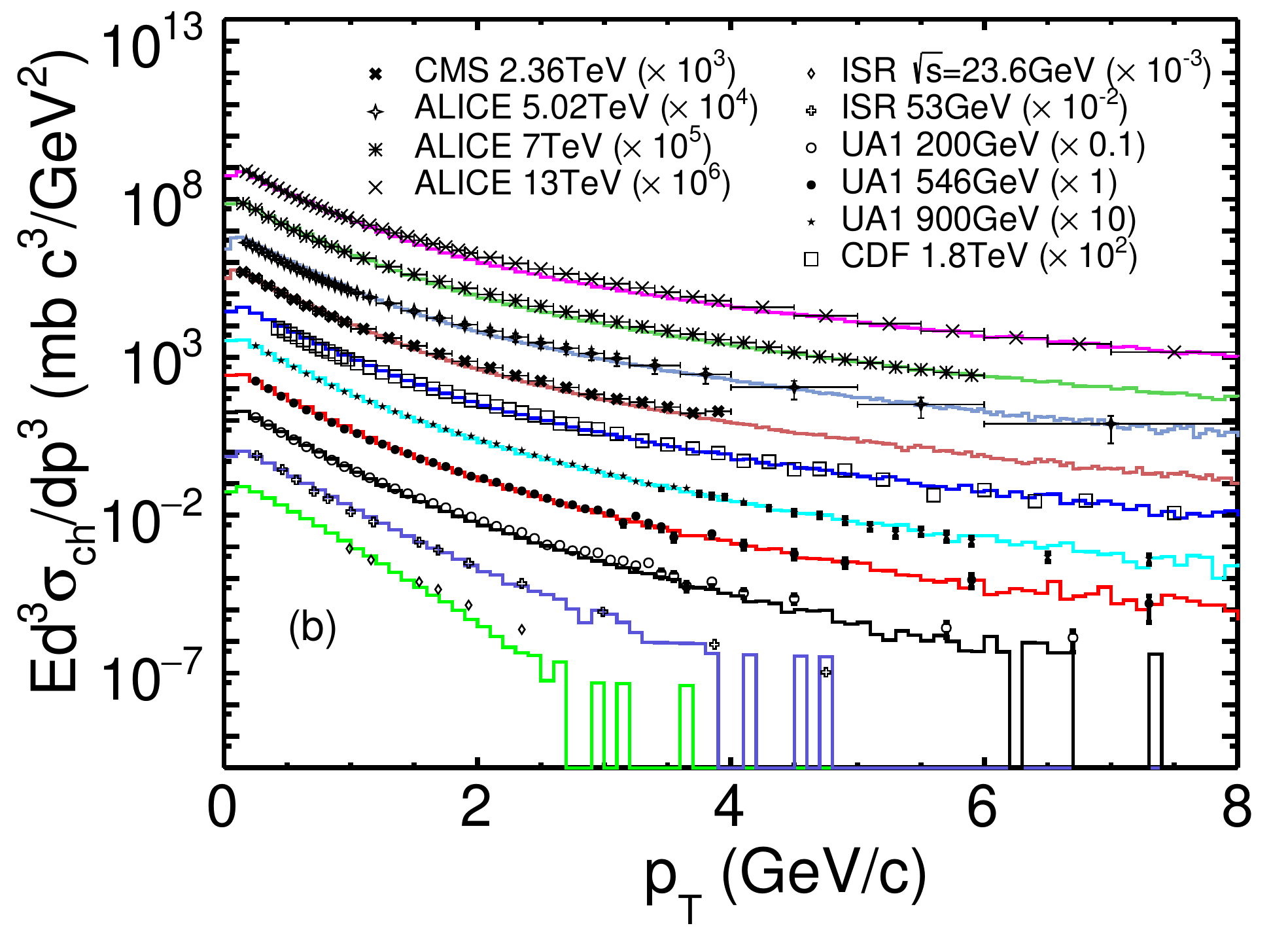}
\caption{(a) Pseudorapidity distributions of charged particles in 
inelastic $pp$ collisions at 23.6 and 53 GeV, NSD $p\bar{p}$ collisions at 200, 546, 900 and 1800GeV, and NSD $pp$ collisions at 2.76, 7 and 13 TeV from the AMPT model in comparison with the experimental data. 
(b) Invariant transverse momentum cross sections around mid-pseudorapidity
from the AMPT model in comparison with the experimental data that also include $pp$ collisions at 2.36 TeV.}
\label{fig:pp}
\end{figure}

Figures~\ref{fig:pp}(a) and \ref{fig:pp}(b) show respectively the $dN_{\rm ch}/d\eta$ distributions and the $p_{\rm T}$ spectra around mid-pseudorapidity 
of charged particles from the string melting AMPT model (curves) in comparison with the experimental data (symbols) in $pp$ or $p\bar{p}$ collisions over a wide energy range. 
We see that using the constant Lund fragmentation parameters $a_L=0.8$ and   $b_L^{pp}=0.7$ GeV$^{-2}$ allows us to reasonably describe these data.

In the $dN_{\rm ch}/d\eta$ distribution we use the same procedure to select the events from the AMPT model calculations as the experimental data. 
The ISR data are for inelastic $pp$ collisions~\cite{Thome:1977ky}.
The ALICE non-single-diffractive (NSD) data~\cite{Adam:2015gka,Adam:2015pza}
refer to events that have at least one charged particle on each side of the V0 detectors which cover the $\eta$ range of $2.8<\eta<5.1$ and $-3.7<\eta<-1.7$, 
while for the UA5~\cite{Alner:1986xu}, CDF~\cite{Abe:1989td} and CMS~\cite{Khachatryan:2010us} data the detectors cover the range of $2<|\eta|<5.6$, $3.2<|\eta|<5.9$ and $2.9<|\eta|<5.2$, respectively. 
For the transverse momentum spectra, we use the same $\eta$ cut as the experimental data:  $|\eta|<0.35$ for $pp$ collisions at 23.6 and 53 GeV~\cite{Alper:1973nv,Thome:1977ky}, $|\eta|<2.5$ for $p\bar{p}$ collisions at 200, 546 and 900 GeV~\cite{Albajar:1989an}, $|\eta|<1$ for $p\bar{p}$ collisions at 1.8 TeV~\cite{Abe:1988yu}, $0<\eta<0.2$ for $pp$ collisions at 2.36 TeV~\cite{Roland:2010ema}, and $|\eta|<0.8$ for $pp$ collisions at 5.02~\cite{Acharya:2018qsh}, 7 and 13 TeV~\cite{Adam:2015pza}. 
Also, the event selection procedure is the same as that used for Fig.~\ref{fig:pp}(a), while the UA1~\cite{Albajar:1989an} selection criterion is the same as UA5.
For the experimental data at LHC energies shown in Fig.~\ref{fig:pp}(b), we have converted the $Ed^3N/dp^3$ data and AMPT $Ed^3N/dp^3$ results 
to $Ed^3\sigma/dp^3$ with the multiplication factor $\sigma_{inel}$. 
Note that the $b_L^{pp}$ value of 0.7 GeV$^{-2}$ here is different than the earlier value of 0.4 GeV$^{-2}$~\cite{Zhang:2019utb} mostly because we now determine its value from a systematic fit to the $\langle p_{\rm T}\rangle$ data. 
We also need to point out that in the earlier study~\cite{Zhang:2019utb} the AMPT results on the invariant transverse momentum cross sections for the lowest two energies (23.6 and 53 GeV) were  mistakenly divided by a factor of two.

\subsection{Au+Au and Pb+Pb collisions}

\begin{figure}[!htb]
\includegraphics[scale=0.43]{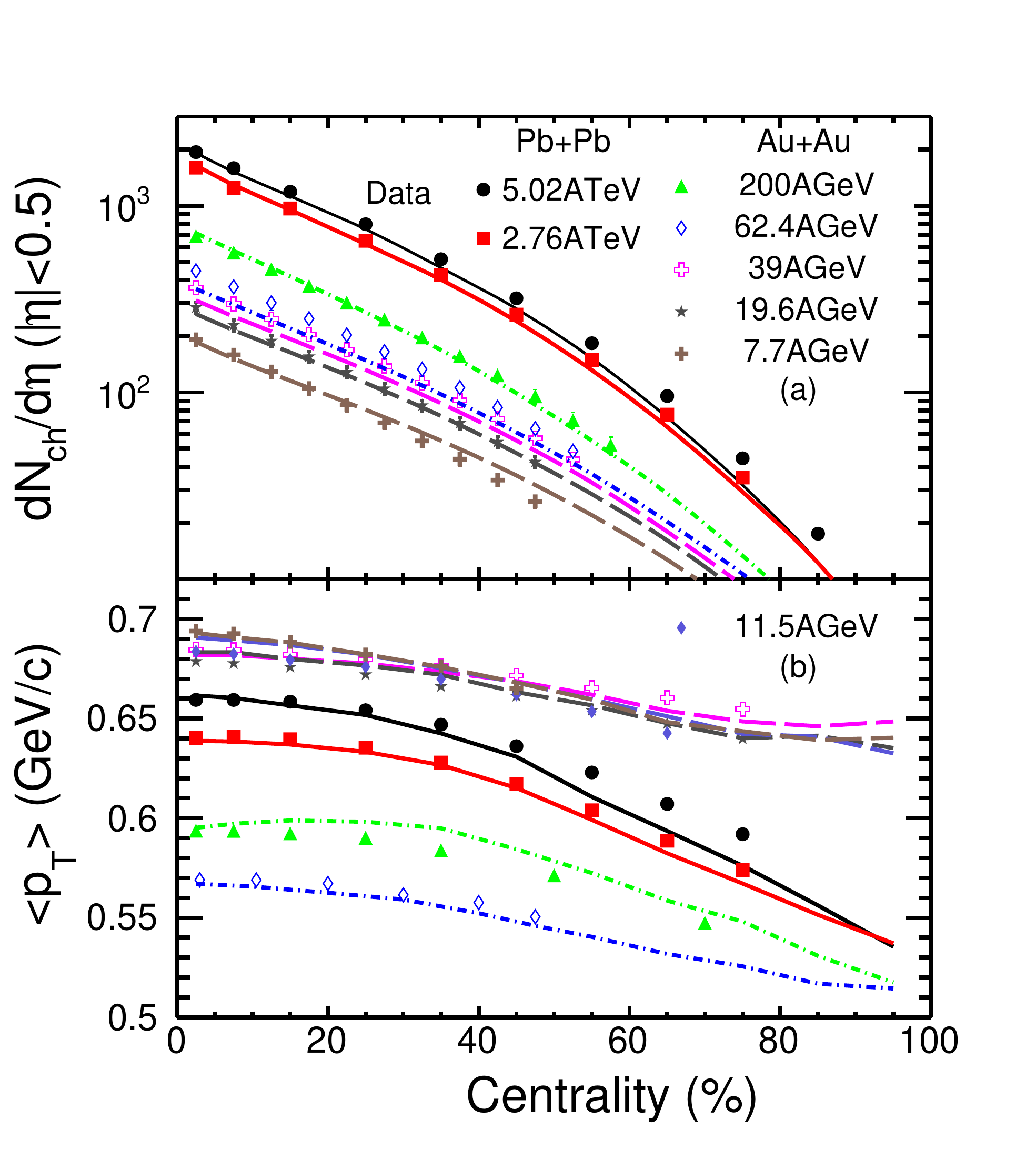}
\caption{$dN_{\rm ch}/d\eta$ within $|\eta|<0.5$ (a) 
and the mean transverse momentum $\langle p_{\rm T}\rangle$ around mid-rapidity (b) versus centrality in Au+Au collisions at RHIC energies and Pb+Pb collisions at LHC energies from the AMPT model (curves) in comparison with the experimental data (symbols). Note the different $p_{\rm T}$ range used for the $\langle p_{\rm T}\rangle$ calculation: [0.15, 2] GeV$/c$ (solid), [0.2, 2] GeV$/c$ (dot-dashed), [0.4, 1.3] GeV$/c$ (dashed).}
\label{fig:aa}
\end{figure}

We now apply the improved AMPT model to Au+Au and Pb+Pb collisions.
Figures~\ref{fig:aa}(a) and \ref{fig:aa}(b)
show respectively the $dN_{\rm ch}/d\eta$ yield at mid-pseudorapidity
and mean transverse momentum $\langle p_{\rm T}\rangle$ around mid-rapidity of charged particles from the AMPT model versus centrality in comparison with experimental data 
for Au+Au collisions at RHIC energies and Pb+Pb collisions at LHC energies. 
We use the same method to determine centrality as the experiments. 
For example, the centrality for the LHC results (from the ALICE Collaboration) is based on the multiplicity of charged particles within $2.8<\eta<5.1$ and $-3.7<\eta<-1.7$, while for the PHENIX, PHOBOS and STAR experiments at RHIC energies the centrality is based on the charged particle multiplicity within $3.0<|\eta|<3.9$, $|\eta|<3.2$, and $|\eta|<0.5$, respectively. 
Note that the $\langle p_{\rm T}\rangle$ values from both the AMPT model and experimental data correspond to charged particles within the $p_{\rm T}$ range of 
[0.4, 1.3] GeV$/c$ for collision energies from $7.7A$ to $39A$ GeV~\cite{Adamczyk:2017iwn}, [0.2, 2] GeV$/c$ at $62.4A$~\cite{Back:2004ra} and $200A$ GeV~\cite{Adams:2003kv}, and [0.15, 2] GeV$/c$ at $2.76A$~\cite{Abelev:2012hxa} and $5.02A$ TeV~\cite{Acharya:2018qsh}. 
Also, results in Fig.~\ref{fig:aa}(b) correspond to the (pseudo)rapidity range of 
$|y| \leq 0.1$ at energies from $7.7A$ to $39A$ GeV,
$0.2<\eta <1.4$ at $62.4A$ GeV, $|\eta| \leq 0.5$ at $200A$ GeV, 
and $|\eta| \leq 0.8$ at $2.76A$ and $5.02A$ TeV.

From Fig.~\ref{fig:aa}(a) we see that the improved AMPT model can reasonably reproduce the mid-pseudorapidity $dN_{\rm ch}/d\eta$ data for the most central (0-5\% centrality) collisions at all these energies except for $39A$ GeV and $62.4A$ GeV, where it underestimates the data. We also see that the model can reasonably describe the centrality dependence of $dN_{\rm ch}/d\eta$ in Au+Au collisions at RHIC energies~\cite{Adare:2015bua}, 
while for Pb+Pb collisions at $2.76A$ TeV~\cite{Abelev:2012hxa} and $5.02A$ TeV~\cite{Adam:2016ddh} it underestimates the $dN_{\rm ch}/d\eta$ for peripheral collisions.
Figure~\ref{fig:aa}(b) shows that the string melting version of the AMPT model describes the energy dependence of $\langle p_{\rm T}\rangle$ reasonably well for Au+Au and Pb+Pb collisions over the colliding energies from $7.7A$ GeV to $5.02A$ TeV. 
The model underestimates the $\langle p_{\rm T}\rangle$ for peripheral collisions at the LHC energies while overestimates the $\langle p_{\rm T}\rangle$ for semi-peripheral and peripheral collisions at the top RHIC energy of $200A$ GeV; however, the difference from the data is no more than $\approx 3\%$.

\begin{figure}[!htb]
\includegraphics[scale=0.43]{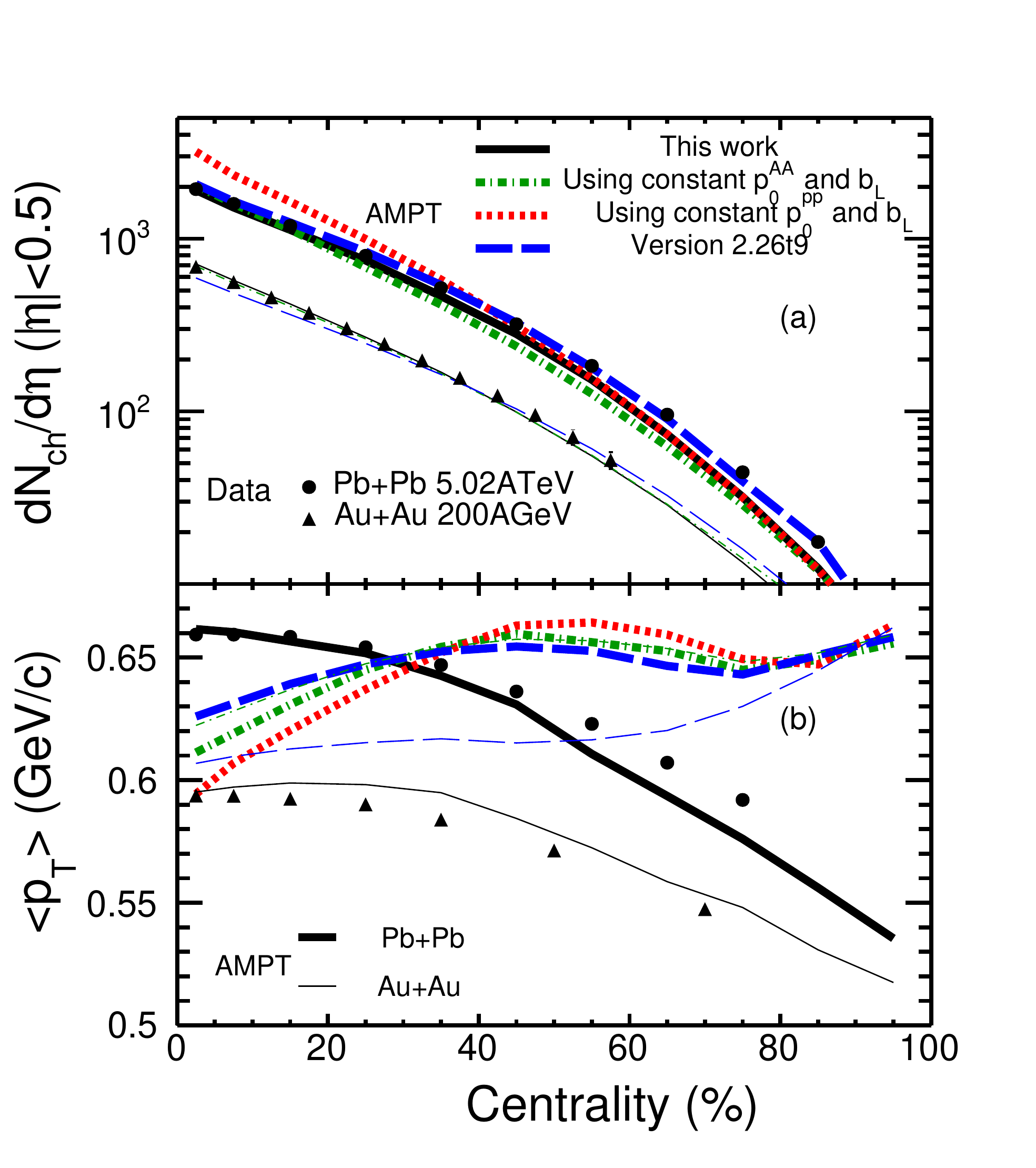}
\caption{$dN_{\rm ch}/d\eta$ within $|\eta|<0.5$ (a) and $\langle p_{\rm T}\rangle$ around mid-pseudorapidity (b) versus centrality in $5.02A$ TeV Pb+Pb  collisions (thick curves) and $200A$ GeV Au+Au collisions (thin curves) from this work (solid curves) and earlier versions of the AMPT model in comparison with the experimental data (symbols); the $p_{\rm T}$ range used for the $\langle p_{\rm T}\rangle$ calculation is [0.15, 2] GeV$/c$ at $5.02A$ TeV and [0.2, 2] GeV$/c$ at $200A$ GeV.}
\label{fig:compare}
\end{figure}

We now compare this work with two earlier versions of the string melting AMPT model 
in Fig.~\ref{fig:compare}(a) for $dN_{\rm ch}/d\eta$ within $|\eta|<0.5$ and in Fig.~\ref{fig:compare}(b) for the $\langle p_{\rm T}\rangle$ around mid-rapidity versus centrality in Pb+Pb collisions at $5.02A$ TeV and Au+Au collisions at $200A$ GeV. 
When we do not use the local nuclear scaling of $p_0$ and $b_L$ but instead use constant $b_L=0.15$ GeV$^{-2}$ and a constant $p_0(s)$ at a given energy for the AMPT model of this work, the model is the same as the one developed in Ref.~\cite{Zheng:2019alz}, 
and we obtain the dot-dashed curves when using $p_0(s)=p_0^{AA}(s)$ and the dotted curves when using $p_0(s)=p_0^{pp}(s)$ (note however that $p_0^{AA}(s)=p_0^{pp}(s)$ at $200A$ GeV).  
Results from the public AMPT version 2.26t9~\cite{note} are also shown (dashed curves) for comparison, where the Lund parameters are taken as $a_L=0.55$ at $200A$ GeV and  0.30 at $5.02A$ TeV with $b_L=0.15$ GeV$^{-2}$~\cite{Lin:2014tya}. 

In Fig.~\ref{fig:compare}(a) we see that the charged particle yield in central Pb+Pb collisions at  $5.02A$ TeV from using $p_0(s)=p_0^{pp}(s)$ is much higher than the experimental data, and it is necessary to use the global nuclear scaling~\cite{Zheng:2019alz}, i.e., $p_0(s)=p_0^{AA}(s)$, to reduce the total minijet cross section and consequently the particle yield. For peripheral collisions, however, the effect from the nuclear scaling of $p_0$ is much smaller because the binary scaling of minijet productions makes $p_0$ less important than for central collisions. These features are essentially the same as our earlier results (Fig.~11 of Ref.~\cite{Zhang:2019utb}). 
We also see as expected that the $dN_{\rm ch}/d\eta$ results from this work are close to the AMPT results using the constant $p_0^{AA}$ for central collisions but close to the AMPT results using the constant $p_0^{pp}$ for peripheral collisions. 
In addition, we see that, compared to the $dN_{\rm ch}/d\eta$ results from the AMPT version 2.26t9, results from this work are slightly worse at $5.02A$ TeV but slightly better at $200A$ GeV.

Previously we found that the centrality dependence of charged particle $\langle p_{\rm T}\rangle$ from the AMPT model is inconsistent with the experimental data at RHIC and LHC~\cite{Ma:2016fve}. This is the case in Fig.~\ref{fig:compare}(b) 
for the results from the AMPT version 2.26t9 (dashed curves). 
Similarly, the AMPT model when using constant $b_L=0.15$ GeV$^{-2}$ and constant $p_0(s)$ (at a given energy)~\cite{Zheng:2019alz} gives the wrong centrality dependence of $\langle p_{\rm T}\rangle$ around mid-pseudorapidity, where the model results (dot-dashed or dotted) show a mostly increasing trend with the increase of centrality while the data show a mostly decreasing trend. 
We also find that the decrease of $\langle p_{\rm T}\rangle$ towards the most central events
from both earlier AMPT versions (dotted, dashed, and dot-dashed curves) is mainly a result of the stronger parton rescatterings in more central collisions.
On the other hand, the local nuclear scaling of this work enables the string melting AMPT model (solid curves) to reasonably reproduce the centrality dependence of charged particle $\langle p_{\rm T}\rangle$ for the first time.

\subsection{Smaller systems including $p$Pb collisions}

\begin{figure}[!htb]
\includegraphics[scale=0.43]{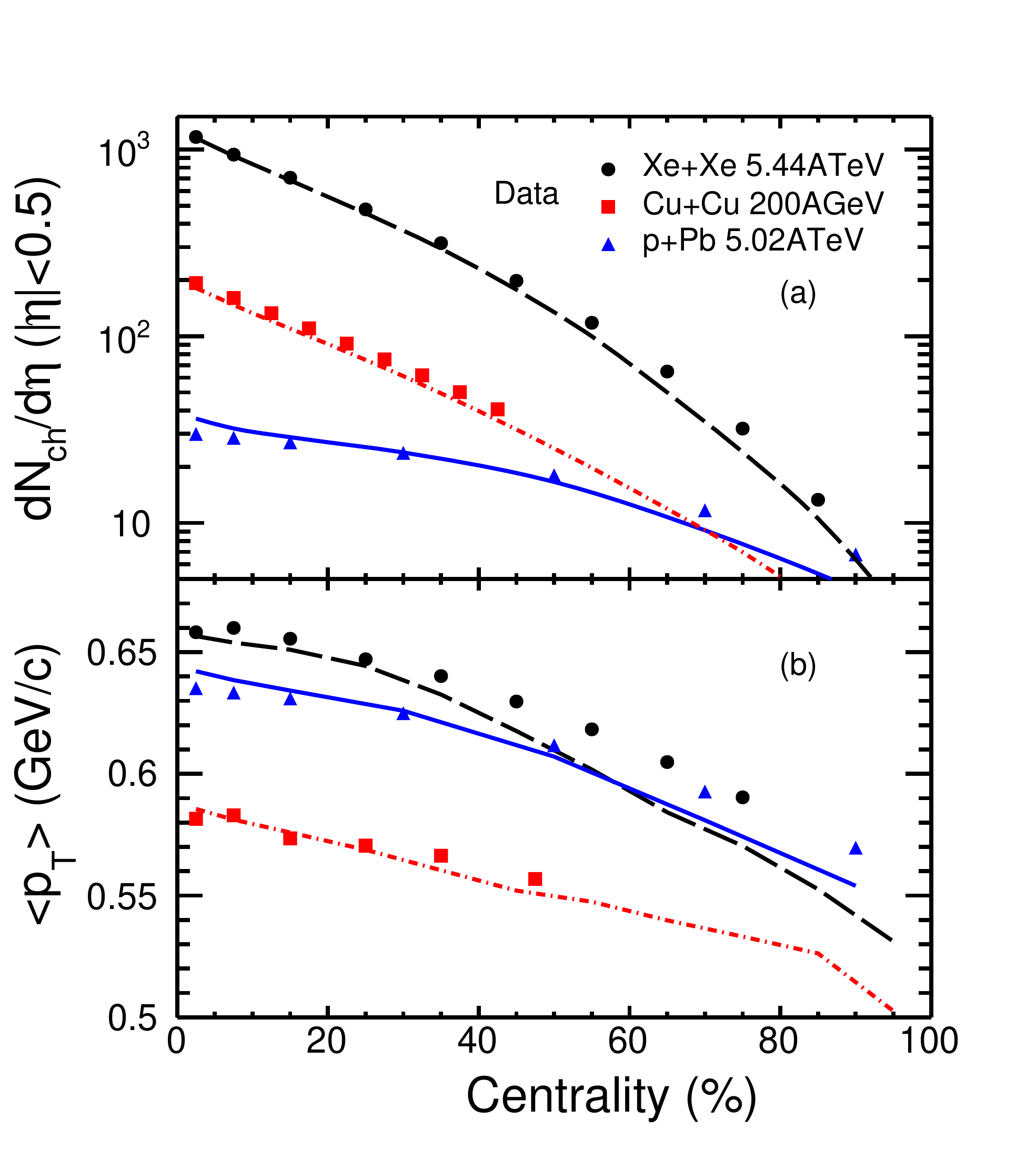}
\caption{$dN_{\rm ch}/d\eta$ within $|\eta|<0.5$ (a) and $\langle p_{\rm T}\rangle$ 
around mid-pseudorapidity (b) versus centrality in Xe+Xe collisions at $5.44A$ TeV, Cu+Cu collisions at $200A$ GeV, and $p$Pb collisions at $5.02A$ TeV from the AMPT model (curves) in comparison with the experimental data (symbols). 
The $p_{\rm T}$ range used for the $\langle p_{\rm T}\rangle$ calculation is [0.2, 2] GeV$/c$ at $200A$ GeV and [0.15, 2] GeV$/c$ at the other two LHC energies.}
\label{fig:small}
\end{figure}

For the system size dependence, it is of particular interest to study the same observables in smaller systems like $pA$ and other $AA$ collisions. 
Figure~\ref{fig:small} shows the results for three smaller collision systems: 
Xe+Xe collisions at $5.44A$ TeV~\cite{Acharya:2018eaq,Acharya:2018hhy}, 
Cu+Cu collisions at $200A$ GeV~\cite{Alver:2005nb,Adare:2015bua}, 
and $p$Pb collisions at $5.02A$ TeV~\cite{Adam:2014qja}. 
We use the same centrality estimator as the experiments, which is the charged particle multiplicity within $2.8<\eta<5.1$ and $-3.7<\eta<-1.7$ for Xe+Xe collisions~\cite{Acharya:2018hhy} and within $|\eta|<3.2$ for Cu+Cu collisions~\cite{Adare:2015bua}.
For $p$Pb collisions at $5.02A$ TeV, the experiment used the energy deposit in the ZDC detector coupled with a heuristic model related to the number of binary collisions ($N_{\rm coll}$) to determine the centrality; due to the lack of slow nucleon physics in the AMPT model we use the model $N_{\rm coll}$ as the centrality estimator in the AMPT model calculations.

Figures~\ref{fig:small}(a) and \ref{fig:small}(b)
show respectively the mid-pseudorapidity $dN_{\rm ch}/d\eta$ 
and $\langle p_{\rm T}\rangle$ of charged particles from the AMPT model versus centrality in comparison with the experimental data for the three collision systems.
The $\langle p_{\rm T}\rangle$ values are calculated for hadrons around mid-pseudorapidity: 
$|\eta|<0.8$ for XeXe collisions, $0.2<\eta<1.4$ for CuCu collisions, and $|\eta|<0.3$ for $p$Pb collisions. We see that the improved AMPT model describes these data rather well,  
confirming the validity of our method of using local nuclear scaling for the $p_0$ and $b_L$ parameters. This is noteworthy because the data of these smaller systems are not considered in the fitting of the parameter functions such as $\alpha(s)$ and $\beta(s)$  
in this study, although the mid-pseudorapidity $dN_{\rm ch}/d\eta$ and $\langle p_{\rm T}\rangle$ data for the most central Au+Au/Pb+Pb collisions have been used. 
Also note that the AMPT results in Fig.~\ref{fig:small}
underestimate both the mid-pseudorapidity $dN_{\rm ch}/d\eta$ and $\langle p_{\rm T}\rangle$ 
for peripheral Xe+Xe collisions; however, in this study we have not included the 
non-spherical deformation of the Xe nucleus~\cite{Moller:2015fba}.

\section{discussions}
\label{sec:discussions}

The local nuclear scalings of the $b_L$ parameter in Eq.\eqref{eq:bl} and $p_0$ parameter in Eq.\eqref{eq:p0} both depend on the geometric mean of the two nuclear thickness functions, $\sqrt{T_A(s_A)T_B(s_B)}$; therefore, this geometric form of scaling is similar to the binary scaling in heavy ion collisions. 
On the other hand, one could also scale the two parameters according the arithmetic mean of the two thickness functions as the following:
\begin{eqnarray}
b_L(s_A,s_B,s)&=&b_L^{pp}/ {\left [ \frac {T_A(s_A)+T_B(s_B)}{2~T_p} \right ]^{\beta(s)}},
\label{eq:blplus} \\
p_0(s_A,s_B,s)&=&p_0^{pp}(s) *\left [ \frac {T_A(s_A)+T_B(s_B)}{2~T_p} \right ]^{\alpha(s)}, 
\label{eq:p0plus}
\end{eqnarray}
and this arithmetic form of local scaling is similar to the participant scaling. 
For symmetric ($AA$) collision systems at impact parameter $b=0$ fm, 
the two different forms are almost identical, because $T_A(s_A)=T_B(s_B)$ is approximately true
which then reduces Eq.~\eqref{eq:blplus} to Eq.\eqref{eq:bl} and Eq.~\eqref{eq:p0plus} to Eq.\eqref{eq:p0}. Therefore, we expect that the different forms will not affect the model results for the most central $AA$ collisions. 
On the other hand, the centrality dependence and the system size dependence of observables could be different for the two different forms. One can expect from Eqs.~\eqref{eq:bl}, \eqref{eq:p0}, \eqref{eq:blplus} and \eqref{eq:p0plus} that the difference between the two forms will be the biggest for the most asymmetric collisions, i.e., central $pA$ collisions. 

\begin{figure}[!htb]
\includegraphics[scale=0.42]{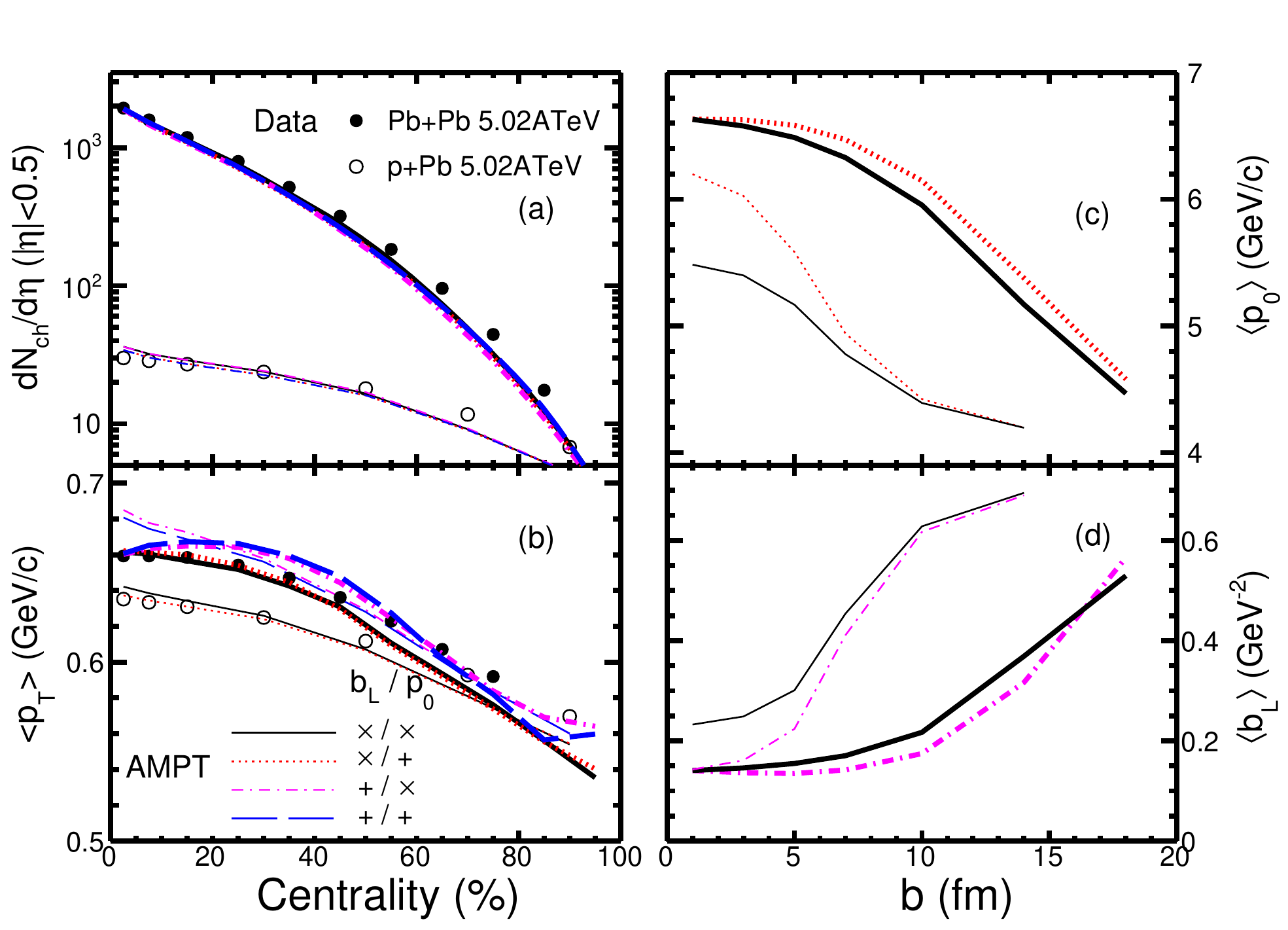}
\caption{$dN_{\rm ch}/d\eta$ within $|\eta|<0.5$ (a) and $\langle p_{\rm T}\rangle$ 
around mid-pseudorapidity (b) versus centrality, as well as $\langle p_0 \rangle$ (c) and $\langle b_L \rangle$ (d) versus the impact parameter, for $5.02A$ TeV Pb+Pb (thick curves) and $p$Pb collisions (thin curves) from different forms of local nuclear scaling in the AMPT model (see text for details); symbols represent the experimental data.}
\label{fig:four-forms}
\end{figure}

In Figs.~\ref{fig:four-forms}(a) and \ref{fig:four-forms}(b) we compare the AMPT model results 
of $dN_{\rm ch}/d\eta$ and $\langle p_{\rm T}\rangle$ around mid-pseudorapidity versus centrality for both Pb+Pb and $p$Pb collisions at $5.02A$ TeV. 
Since one can choose separate forms for the local nuclear scaling of $b_L$ and $p_0$, the model results including four different combinations, e.g., the curves labeled as ``x/x'' represent our default results of using the geometric form for both $b_L$ and  $p_0$, while the curves labeled as ``x/+'' represent the model results of using the geometric form
of Eq.~\eqref{eq:bl} for $b_L$ but the arithmetic form of Eq.~\eqref{eq:p0plus} for $p_0$. 
We see that different forms have a relatively small effect on the $dN_{\rm ch}/d\eta$ yield and its the centrality dependence.
On the other hand, they have a significant effect on the mean transverse momentum, especially for $p$Pb collisions, where the arithmetic form for the $b_L$ scaling significantly over-predicts the experimental data, regardless of the form used for the $p_0$ scaling.
Therefore, we choose the geometric form for the local scalings of both $b_L$ and $p_0$ as shown in Eqs.~\eqref{eq:bl} and \eqref{eq:p0}, while we note that the arithmetic form for the scaling of $p_0$ would work similarly well as indicated by the dotted curves in Fig.~\ref{fig:four-forms}. We note that a Bayesian analysis of the TRENTo initial condition~\cite{Moreland:2014oya} with a hybrid model found that the geometric form for the initial state entropy deposition is preferred by the experimental data than several other forms including the arithmetic form~\cite{Bernhard:2016tnd}. 

We show in Figs.~\ref{fig:four-forms}(c) and \ref{fig:four-forms}(d) the impact parameter dependence of $\langle p_0 \rangle$ and $\langle b_L \rangle$, respectively, 
from the two different forms of local nuclear scaling. 
Indeed, the difference between the geometric and arithmetic forms is the biggest for central $p$Pb collisions, where the arithmetic form gives a bigger variation of the $\langle p_0 \rangle$ and $\langle b_L \rangle$ values with the impact parameter. 
As a result, the higher $p_0$ value from the arithmetic form leads to a lower $dN_{\rm ch}/d\eta$ while the lower $b_L$ value from the arithmetic form gives a higher 
$\langle p_{\rm T}\rangle$ for $p$Pb collisions. 

\section{summary}
\label{sec:summary}

A multi-phase transport model can describe multiple observables in relativistic heavy ion collisions and can thus be very useful for the study of the dynamics and physical properties of the dense matter. 
However, certain key parameters need to have significantly different values for $pp$ and central $AA$ collisions for the model to well describe the yield and transverse momentum spectrum of the bulk matter. 
In this study we use local nuclear scaling to relate two key parameters in the initial condition to the nuclear thickness functions of the two colliding nuclei so that the parameter values change with the system size self consistently. 
Specifically, we let two parameters in the string melting AMPT model with modern parton distribution functions of nuclei, the Lund string fragmentation parameter $b_L$ and the minijet transverse  momentum cutoff $p_0$, to scale with powers of $\sqrt {T_A(s_A) T_B(s_B)}$ similar to the number of binary collisions. 
We then systematically study charged particle productions in different collision systems over a wide energy range.

We start from the parameter values for $pp$ collisions that allow a good description of the charged particle yields and transverse momentum spectra in $pp$ collisions from 23.6 GeV to 13 TeV. We then determine the two energy-dependent power functions in the local nuclear scaling of the $p_0$ and $b_L$ parameters by comparing to data on the charged particle $dN_{\rm ch}/d\eta$ and mean transverse momentum $\langle p_{\rm T}\rangle$ (below $p_{\rm T} \lesssim 2$  GeV) around mid-pseudorapidity in the most central Au+Au and Pb+Pb collisions. Then the centrality dependence and system size dependence are model predictions. 
We show that, for charged particles around mid-pseudorapidity in Au+Au collisions from $7.7A$ GeV to $200A$ GeV and Pb+Pb collisions at LHC energies, the improved AMPT model not only provides reasonable descriptions of the centrality dependence of the $dN_{\rm ch}/d\eta$ yield but also for the first time well describes the centrality dependence of $\langle p_{\rm T} \rangle$. The model also provides reasonable descriptions of smaller systems including $p$Pb, Cu+Cu and Xe+Xe collisions without any change of the parameter functions. 
This work allows a multi-phase transport model to describe the system size and centrality dependences of nuclear collisions self consistently, making the model more reliable for further studies of nuclear collisions from small to large systems.

\begin{acknowledgments}
This work is supported by the National Natural Science Foundation of China under Grants No. 11890711 (C.Z. and S.S.S.) and No. 11905188 (L.Z.), the National Key Research and Development Program of China under Grant No. 2020YFE0202002 (C.Z. and S.S.S.), the Chinese Scholarship Council (C.Z.), and the National Science Foundation under Grant No. 2012947 (Z.-W.L.).

\end{acknowledgments}

\end{document}